\documentclass[twocolumn,howpacs,preprintnumbers,amsmath,amssymb,aps]{revtex4}

\usepackage{amssymb}
\usepackage{amsmath,amssymb,graphicx}
\usepackage{graphicx}
\usepackage{dcolumn}
\usepackage{bm}
\usepackage{axodraw}
\usepackage{epsfig}

\def\be{\begin{equation}}
\def\ee{\end{equation}}
\def\bea{\begin{eqnarray}}
\def\eea{\end{eqnarray}}

\def\nn{\nonumber}

\newcommand{\bers}{\begin{eqnarray*}}
\newcommand{\eers}{\end{eqnarray*}}
\newcommand{\bt}{\begin{itemize}}
\newcommand{\et}{\end{itemize}}
\def\lsim{\raise0.3ex\hbox{$\;<$\kern-0.75em\raise-1.1ex\hbox{$\sim\;$}}}
\def\gsim{\raise0.3ex\hbox{$\;>$\kern-0.75em\raise-1.1ex\hbox{$\sim\;$}}}

\def\ie{{\it i.e.}}

\begin{document}
\preprint{}

\title{TeV Scale Gauged $B-L$ With Inverse Seesaw Mechanism}
\author{Shaaban Khalil}
\affiliation{Center for Theoretical Physics at the
British University in Egypt, Sherouk City, Cairo 11837, Egypt.\\
Department of Mathematics, Ain Shams University, Faculty of
Science, Cairo, 11566, Egypt.}
\date{\today}

\begin{abstract}
We propose a modified version of the TeV scale $B-L$ extension of
the standard model, where neutrino masses are generated through the
inverse seesaw mechanism. We show that heavy neutrinos associated
with this model can be accessible via clean signals at the LHC. The
search for the extra gauge boson $Z'_{B-L}$ through
the decay into dileptons or two dileptons plus missing energy is
studied. We also show that the $B-L$ extra Higgs can be directly probed
at the LHC via a clean dilepton and missing energy signal.

\end{abstract}
\maketitle
%
%
%
The search for new physics at TeV scale is a major goal of the
Large Hadron Collider (LHC). Non-vanishing neutrino masses
represent a firm observational evidence of new physics beyond the
standard model (SM). TeV scale Baryon minus Lepton (B-L) extension
of the SM, which is based on the gauge group $SU(3)_C \times
SU(2)_L \times U(1)_Y \times U(1)_{B-L}$, has been recently
proposed \cite{Khalil:2006yi} as the simplest model beyond the SM
that provides a viable and testable solution to the neutrino mass
mystery of contemporary particle physics. There have been several
attempts in the past to extend the gauge symmetry of the SM with
$U(1)_{B-L}$, see for example Ref.\cite{Buchmuller:1991ce}.

In this model, three SM singlet fermions arise quite naturally due
to the anomaly cancellation conditions. These three particles are
accounted for right handed neutrinos, and hence a natural
explanation for the seesaw mechanism is obtained. In addition, the
model also contains an extra gauge boson corresponding to $B-L$
gauge symmetry and an extra SM singlet scalar (heavy Higgs). If
the scale of $B-L$ breaking is of order TeV, these new particles
will lead to very interesting signatures at the LHC
\cite{Emam:2007dy,Basso:2009hf,Huitu:2008gf,Iso:2009nw}. In
general, the scale of $B- L$ symmetry breaking is unknown, ranging
from TeV to much higher scales. However, it was proven
\cite{Khalil:2007dr} that in supersymmetric framework, the scale
of B-L is nicely correlated with the soft supersymmetry breaking
scale, which is TeV. Recently, there has been considerable
interest in studying the phenomenological implications of TeV
$B-L$ model at colliders
\cite{Basso:2009hf,Emam:2008zz,Iso:2009nw}.

In TeV scale $B-L$ extension of the SM, the Majorana neutrino
Yukawa interaction: $\lambda_{\nu_R} \chi \bar{\nu}_R^c \nu_R$
induces the following masses for the right-handed neutrinos after
$U(1)_{B-L}$ symmetry breaking: $M_{\nu_R} = \lambda_{\nu_{R}}
v'$, where $v' =\langle \chi \rangle$ is the vacuum expectation
value (vev) of the $B-L$ symmetry breaking. Below the Electroweak
(EW) symmetry breaking, Dirac neutrino masses, $m_D =
\lambda_{\nu} v$, are generated. Here $v$ is the vev of the EW
symmetry breaking and $\lambda_{\nu}$ are the Dirac neutrino
Yukawa couplings. Therefore, the physical light neutrino masses
are given by $m^2_D/M_{\nu_R}$, which can account for the measured
experimental results if $\lambda_{\nu}\lsim 10^{-6}$. Such small
couplings may be considered as unnatural fine-tuning.
Nevertheless, they induce new interaction terms between the heavy
neutrino, weak gauge boson $W$ and $Z$, and the associated
leptons. These couplings play important role in the decay of
lightest heavy neutrino at the LHC
\cite{Abbas:2007ag,Huitu:2008gf}. This signal is one of the
striking signatures of TeV scale $B-L$ extension of the SM.

It is very important to note that the above analysis, which led to
severe constraints on the neutrino Yukawa couplings, were based on
the canonical type-I seesaw mechanism. In this paper, we propose a
new modification for our TeV scale $B-L$ model
\cite{Khalil:2006yi}, to prohibit type I seesaw and allow another
scenario for generating light neutrino masses, namely the inverse
seesaw mechanism \cite{Mohapatra:1986bd,GonzalezGarcia:1988rw}.
Our modification is based on the following: $(i)$ The SM singlet
Higgs, which breaks the $B-L$ gauge symmetry, has $B-L$ charge
$=-1$. $(ii)$ The SM singlet fermion sector includes two singlet
fermions with $B-L$ charges $=\pm 2$ with opposite matter parity.
In this case, we will show that small neutrino masses can be
generated through the inverse seesaw mechanism, without any
stringent constraints on the neutrino Yukawa couplings. Therefore,
a significant enhancement of the verifiability of TeV scale $B-L$
extension of the SM is obtained.

The proposed TeV scale $B-L$ extension of the SM is based on the
gauge group $SU(3)_C\times SU(2)_L\times U(1)_Y\times U(1)_{B-L}$,
where the $U(1)_{B-L}$ is spontaneously broken by a SM singlet
scalar $\chi$ with $B-L$ charge $=-1$. As in the previous model, a
gauge boson $Z'_{B-L}$ and three SM singlet fermions $\nu_{R_i}$
with $B-L$ charge $=-1$ are introduced for the consistency of the
model. Finally, three SM singlet fermions $S_1$ with $B-L$ charge
$=-2$ and three singlet fermions $S_2$ with $B-L$ charge $=+2$ are
considered to implement the inverse seesaw mechanism.

The Lagrangian of the leptonic sector in this model is given by%
\bea%
{\cal L}_{B-L}&=&-\frac{1}{4} F'_{\mu\nu}F'^{\mu\nu} + i~ \bar{L}
D_{\mu} \gamma^{\mu} L + i~ \bar{e}_R D_{\mu} \gamma^{\mu} e_R \nonumber\\
&+& i~ \bar{\nu}_R D_{\mu} \gamma^{\mu} \nu_R + i~
\bar{S}_1 D_{\mu} \gamma^{\mu} S_1 + i~ \bar{S}_2 D_{\mu} \gamma^{\mu} S_2\nonumber\\
&+& (D^{\mu} \phi)^\dag(D_{\mu} \phi) + (D^{\mu} \chi)^\dag(D_{\mu} \chi)-
V(\phi, \chi)\nonumber\\
&-& \Big(\lambda_e \bar{L} \phi e_R + \lambda_{\nu} \bar{L}
\tilde{\phi} \nu_R +
\lambda_{S} \bar{\nu}^c_R \chi S_2 + h.c.\Big)\nonumber\\
&-& \frac{1}{M^3}\bar{S}^c_{1} {\chi^\dag}^{4} S_{1}-\frac{1}{M^3}\bar{S}^c_{2} {\chi}^{4} S_{2},%
\label{lagrangian}
\eea %
where $F'_{\mu\nu} = \partial_{\mu} Z'_{\nu} - \partial_{\nu}
Z'_{\mu}$ is the field strength of the $U(1)_{B-L}$. The covariant
derivative $D_{\mu}$ is generalized by adding the term $i g^{''}
Y_{B-L} Z'_{\mu}$, where $g^{''}$ is the $U(1)_{B-L}$ gauge
coupling constant and $Y_{B-L}$ is the $B-L$ quantum numbers of
involved particles. Since $U(1)_{B-L}$ is not orthogonal to
$U(1)_Y$, a mixing term between the two field strengths is
expected. However, in the basis of diagonalizing kinetic terms,
one finds $i g^{''} Y_{B-L} \to i (\tilde{g} Y +  g^{''}
Y_{B-L})$, where $\tilde{g}$ is parameterizing the mixing between
the neutral gauge bosons: $Z$ and $Z'$, which is constrained
experimentally to be small. Therefore, setting $\tilde{g}=0$ is an
acceptable approximation. In this case, what is called minimal
$B-L$ model is obtained \cite{Khalil:2006yi,Basso:2009hf}.
Furthermore, in order to prohibit a possible large mass term $M
S_1 S_2$ in the above Lagrangian, we assume that the SM particles,
$\nu_R$, $\chi$, and $S_2$ are even under matter parity, while
$S_1$ is an odd particle. Finally, $V(\phi, \chi)$ is the most
general Higgs potential invariant under these symmetries and can
be found in Ref.\cite{Khalil:2006yi}.

The non-vanishing vacuum expectation value (vev) of $\chi$:
$|\langle\chi\rangle|= v'/\sqrt 2$ is assumed to be of order TeV,
consistent with the result of radiative $B-L$ symmetry breaking
found in gauged $B-L$ model with supersymmetry
\cite{Khalil:2007dr}. The vev of the Higgs field $\phi$:
$|\langle\phi^0\rangle|= v/\sqrt 2$ breaks the EW symmetry, \ie,
$v =246$ GeV. After the $B-L$ gauge symmetry breaking, the gauge
field $Z^{\prime}$ acquires the following mass: $
M_{Z^{\prime}_{B-L}}^2=g''^2 v'^2$. The bound on $B-L$ gauge
boson, due to negative search at LEP II, implies that $
M_{Z'_{B-L}} /g'' > 6 ~ \rm{TeV}$. This indicates that $v' \gsim
O(\rm{TeV})$. If the coupling $g'' < O(1)$, then one obtains
$m_{Z^{\prime}} \gsim O(600)$ GeV.
Now, we turn to neutrino masses in this model. As can be seen from
Eq.(\ref{lagrangian}), after $B-L$ and EW symmetry breaking, the
neutrino Yukawa interaction terms lead to the following mass
terms:%
\be%
{\cal L}_m^{\nu} = m_D \bar{\nu}_L \nu_R + M_N \bar{\nu}^c_R S_2+h.c.,%
\ee
where $m_D=\frac{1}{\sqrt{2}}\lambda_\nu v$ and $ M_N =
\frac{1}{\sqrt 2}\lambda_{\nu_R} v' $. From this Lagrangian, one
can easily observe that although the lepton number is broken
through the spontaneous $B-L$ symmetry breaking, a remnant
symmetry: $(-1)^{L+S}$ is survived, where $L$ is the lepton number
and $S$ is the spin. After this global symmetry is broken at much
lower scale, a mass term for $S_2$ (and possibly for $S_1$ as
well) is
generated. Therefore, the Lagrangian of neutrino masses, in the flavor basis, is given by: %
\be%
{\cal L}_m^{\nu} =\mu_s \bar{S}^c_2 S_2 +(m_D \bar{\nu}_L \nu_R + M_N \bar{\nu}^c_R S_2 +h.c.) ,%
\ee%
where $\mu_s=\frac{v'^4}{4 M^3}\sim10^{-9}$. The possibility of
generating small $\mu_s$ radiatively, in general inverse seesaw
model, has been discussed in Ref. \cite{Ma:2009gu}.

In the basis $\{\nu^c_L,\nu_R, S_2\}$, the $9\times 9$ neutrino
mass
matrix takes the form:%
\be
\left(%
\begin{array}{ccc}
  0 & m_D & 0\\
  m_D^T & 0 & M_N \\
  0 & M_N^T & \mu_s\\
\end{array}%
\right). %
\ee%
The diagonalization of this mass matrix
leads to the following light and heavy neutrino masses respectively: %
\bea%
m_{\nu_l} &=& m_D M_N^{-1} \mu_s (M_N^T)^{-1} m_D^T,\\
m_{\nu_H}^2 &=& m_{\nu_{H'}}^2 = M_N^2 + m_D^2. %
\eea %
Thus, the light neutrino mass can be of order eV, as required by
the oscillation data, for a TeV scale $M_N$, provided $\mu_s$ is
sufficiently small, $\mu_s \ll M_N$. In this case, the Yukawa
coupling $\lambda_\nu$ is no longer restricted to a very small
value and it can be of order one. Therefore, the possibility of
testing this type of model in LHC is quite feasible.

In general, the physical neutrino states are given
in terms of $\nu^c_L$, $\nu_R$, and $S_2$ as follows:%
\bea %
\nu_{l} &=& \nu^c_L  + a_1 ~\nu_R + a_2~ S_2,
\label{nul}\\
\nu_{H} &=&  a_3 ~\nu^c_L + \alpha~ \nu_R - \alpha~
S_2,
\label{nuH}\\
\nu_{H'} &=& \alpha ~\nu_R + \alpha ~S_2.
\label{nuH'}%
\eea %
For $m_D \simeq 100$ GeV, $M_N \simeq 1$ TeV and $\mu_s \simeq 1$
KeV, one finds that $a_{1,2} \sim m_D/(M_N \sqrt{2+ 2m_D/M_N})
\sim {\cal O}(0.05)$, $a_3 \sim m_D/M_N \sim {\cal O}(0.1)$ and
$\alpha \sim \sin\pi/4$. Therefore, one of the heavy neutrinos of
this model can be accessible via a clean signal at LHC, as will be
discussed below.

It is worth mentioning that the light neutrinos $\nu_l$ have
suppressed mixing (of order $m_D \mu_s/(M_N^2 +m_D^2)$) with one
type of the heavy neutrinos (say $\nu_{H'}$) and a large mixing
(of order $m_D/M_N$) with the other type of heavy neutrinos
($\nu_H$). The mixing between the heavy neutrino $\nu_H$ and
$\nu_H'$ is maximal. The heavy neutrinos $\nu_H$ and $\nu_H'$ can
mediate the lepton flavor processes, like $\mu \to e \gamma$. The
$\mu \to e \gamma$ decay mediated by these heavy neutrinos have
branching ratios \cite{Dev:2009aw}:%
\bea%
BR(\mu \to e \gamma) &\simeq& \frac{\alpha_W^3 \sin^2 \theta_W
m_{\mu}^5}{256 \pi^2 M_W^4 \Gamma_{\mu}} \nonumber\\
&\times&\left\vert \sum_{i=1}^3 (a_3)_{\mu i}(a^*_{3})_{e i}
I\left(\frac{m_{\nu_{H_i}^2}}{M_W^2}\right)\right\vert^2\!\!,~ ~%
\eea%
where $\Gamma_{\mu}$ is the total decay width of $\mu$ and the
loop function $I(x)$ can be found in Ref.\cite{Dev:2009aw}. From
the present experimental limit: $BR(\mu \to e \gamma)$ one
finds%
\be%
\left\vert (a_3)_{\mu \mu}(a^*_{3})_{e \mu}
I\left(\frac{m_{\nu_{H_2}^2}}{M_W^2}\right)\right\vert < 10^{-4}.%
\ee%
Thus for $(a_3)_{\mu \mu} \simeq 0.1$, one obtains the following
constraint on the off-diagonal element $(a_3)_{12}$: %
\be %
(a_{3})_{12} \simeq (m_D M_N^{-1})_{12} < 10^{-3}. %
\ee%
It is important to notice that within the inverse seesaw
mechanism, the branching ratio of $l_{\alpha} \to l_{\beta}
\gamma$ are significantly enhanced compared to the results in the
conventional type I seesaw model.

The LHC discovery of $Z'_{B-L}$ is considered as a smoking gun for
TeV scale $B-L$ extension of the SM. In minimal $B-L$ model, it
was shown that $Z' \to l^+ l^-$ gives the dominant decay channel
with $BR(Z' \to l^+ l^-) \simeq 20\%$. Therefore, the search for
$Z'$ can be accessible via a dilepton channel for $600~{\rm GeV}
\leq M_{Z'} \leq 2~{\rm TeV}$. In our new model of $B-L$ with
inverse seesaw, the decay widths of $Z'$ into lightest heavy
neutrinos $\nu_H$ and $\nu_{H'}$ are given by:%
\bea %
\!\!\!\!\Gamma(Z' \to \nu_H \nu_H) \!&\!=\!&\! \frac{(g''
Y^{\nu_H}_{B-L})^2}{48
\pi} M_{Z'}\!\left(1-4\frac{m_{\nu_H}^2}{M_{Z'}^2}\right)^{3/2}\!\!\!\!\!\!\!,\nn\\
\!\!\!\!\Gamma(Z' \to \nu_{H'} \nu_{H'}) \!&\!=\!&\! \frac{(g''
Y^{\nu_{H'}}_{B-L})^2}{48 \pi} M_{Z'}\!
\left(1-4\frac{m_{\nu_{H'}}^2}{M_{Z'}^2}\right)^{3/2}\!\!\!\!\!\!\!\!.~
\eea %
From Eqs. (\ref{nuH},\ref{nuH'}), the charges $Y_{B-L}^{\nu_H}$
and $Y_{B-L}^{\nu_{H'}}$ are given by %
\bea%
Y_{B-L}^{\nu_{H}} \!\!&\!\simeq\!&\!\! a_3^2 Y_{B-L}^{\nu_L}\!+\!
\alpha \left(Y_{B-L}^{\nu_R^c} \!-\! Y_{B-L}^{S_2}\right)
\!\simeq\! 3 \alpha^2 \!\simeq\! \frac{3}{2},\\
Y_{B-L}^{\nu_{H'}} &\simeq& \alpha^2 \left(Y_{B-L}^{\nu_R^c} +
Y_{B-L}^{S_2}\right) = -\alpha^2 \simeq \frac{-1}{2}.
\eea%

\begin{figure}[t]
\hspace{-0.5cm}\epsfig{file=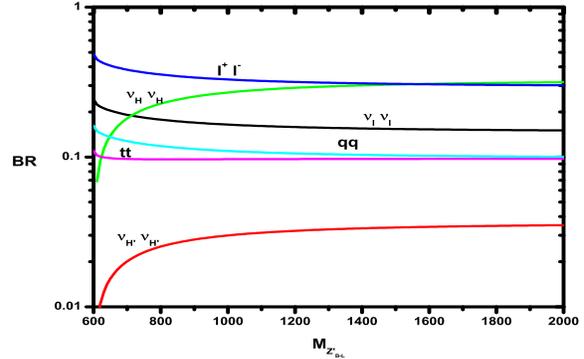, width=8.5cm, height=5.5cm,
angle=0}
\vskip -0.5cm \caption{Branching ratios of $Z'_{B-L}$ as function
of $M_{Z'_{B-L}}$.} \label{fig:BR}
\end{figure}

\begin{figure}[t]
\epsfig{file=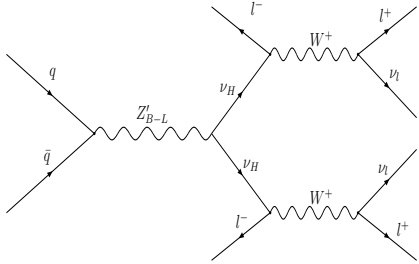, width=5.5cm, height=3.5cm, angle=0}
\caption{$Z'_{B-L}$ production and decay via 2 dilepton plus
missing energy at LHC.} \label{fig2}
\end{figure}

Thus, for heavy $Z'_{B-L}$ ($M_{Z'_{B-L}} \gg 2 M_{\nu_H}$), the
decay channel $Z'_{B-L} \to \nu_H \nu_H$ could be the dominant. In
Fig.\ref{fig:BR} we present the decay branching ratios of
$Z^{\prime } \to f \bar{f}$ as a function of $M_{Z^{\prime }}$ for
$f=l^-, \nu_H, \nu_l, \nu_{H'}, q=u,c,d,s,b$, and $f=t$. As can be
seen from this figure, the decay $Z' \to l^+ l^-$ is the dominant
if $M_{Z'_{B-L}} < 2 M_{\nu_{H}}$. However, for $M_{Z'_{B-L}} \gg
2 M_{\nu_{H}}$, the decay $Z'_{B-L} \to \nu_H \nu_H$ becomes
dominant with branching ratio $\gsim 32\%$. Therefore, searching
for $Z'_{B-L}$ can be easily accessible at the LHC via:
$(i)$ A clean dilepton signal, which can be one of the first new
physics signatures to be observed at the LHC, if $Z'_{B-L}$ is
lighter than twice $\nu_H$ mass. As emphasized in Ref.\cite{Emam:2008zz},
$Z'_{B-L}$ can be discovered in this case, within a mass range
$[800, 1200]$ GeV and an integrated luminosity of $100 ~{\rm
pb}^{-1}$.
$(ii)$ A signal of 2-dilepton plus missing energy, with a tiny SM
background  if $M_{Z'_{B-L}} \gg 2 M_{\nu_H}$. In this case, one
considers the $Z'_{B-L}$ decay into two heavy neutrinos. This
process could enhance the $\nu_H$ production cross section, due to
the resonant contribution from $Z'_{B-L}$ exchange in the
$s$-channel. Then, the $\nu_H$ mainly decays through the $W$ gauge
boson to lepton and neutrino, as shown in Fig.\ref{fig2}. As
explained in Ref.\cite{Huitu:2008gf}, these decays are very clean
with four hard lepton, therefore they are distinctive LHC signals
with nearly free background. Note that in this model, the coupling
of $\nu_H W l$ is of order $0.05  g_2$, which is not very
suppressed as in the minimal $B-L$ model. Therefore, the decay
width of $\nu_H \to W^+ l^-$ is not very small, and hence $\nu_H$
is no longer is a long-lived particle. This could be a distinguish
difference between the two $B-L$ scenarios \cite{Khalil:2009tm}.
%

After the breakdown of the $B-L$ and EW symmetry, mixing between
$\phi$ and $\chi$ is generated. The mixing between the neutral
scalar components of Higgs multiplets, $\phi^0$ and $\chi^0$,
leads to the following mass eigenstates:  SM-like Higgs boson $H$
and heavy Higgs boson $H'$:
\be %
\left(\begin{array}{c} H\\
H^{\prime} \end{array} \right)= \left(\begin{array}{cc} \cos \theta & - \sin\theta\\
\sin\theta & \cos \theta \end{array} \right)  \left(\begin{array}{c} \phi^0\\
\chi^0 \end{array} \right),%
\ee where the mixing angle $\theta$ is defined by %
\be %
\tan 2 \theta = \frac{\vert \lambda_3 \vert v v'}{\lambda_1 v^2 -
\lambda_2 v'^2}. %
\ee %
The masses of $H$ and $H^{\prime}$ are given by
\be%
\!\!m^2_{H,H^{\prime}}\! =\! \lambda_1 v^2 \!+\! \lambda_2 v'^2
\mp
\sqrt{(\lambda_1 v^2 \!-\! \lambda_2 v'^2)^2 \!+\! \lambda_3^2 v^2 v'^2}.%
\ee%
From these expressions, it is clear that $\lambda_3$ is the
measuring of the mixing between the SM Higgs and the $B-L$ extra
Higgs.

As in the minimal $B-L$ model \cite{Emam:2007dy}, the couplings
among the SM-like Higgs, H, and the SM fermions and gauge bosons
are modified by a factor of $\cos\theta$. It is interesting to
note that a maximum mixing with $\theta =\pi/4$ can be obtained if
$\lambda_1 v^2 - \lambda_2 v'^2=0$, which implies that $m_H \simeq
m_{H'}$. However, the restriction from precision EW measurements,
in particular the fit of the parameters $S$, $T$, and $U$,
impose the following constraint on Higgs mixing angle\cite{Dawson:2009yx}: %
\be %
\!{\rm For}~ m_H \!>\! 120 {\rm GeV}~\! \& ~m_{H'}\!>\! 500 {\rm
GeV}
\!\Longrightarrow \!\cos \theta \!>\! 0.9. %
\ee %
Therefore, the cross sections of the SM-like Higgs production
cross sections and decay branching ratios are slightly changed.
Also, the decay widths of $H'$ into SM fermions are suppressed by
$\sin^2 \theta$ factor. Due to a large mixing between light and
heavy neutrinos in this model, the decay channels $H' \to \nu_l
\nu_H$, $H' \to \nu_H \nu_H$ and $H' \to \nu_{H'} \nu_{H'}$ (in
case of $m_{H'} > m_{\nu_H}$, $m_{H'} > 2 m_{\nu_H}$, and $m_{H'}
> 2 m_{\nu_{H'}}$ respectively) are relevant and may lead to important effects.
The decay widths of these channels are given by%
\bea%
\!\!\Gamma(H'\to \nu_l \nu_H)\!&\!=\!&\! \frac{\vert \lambda_S a_2
\vert^2}{32 \pi} m_{H'}  \cos^2 \theta \!\left[1\!-\!
\frac{m_{\nu_H}^2}{m_{H'}^2}
\right]^2\!\!\!,\\
\Gamma(H'\to \nu_H \nu_H)\!&\simeq&\!\Gamma(H'\to \nu_{H'}
\nu_{H'}) \!\simeq \!\Gamma(H'\to \nu_H \nu_{H'}) \nn\\
&\simeq& \frac{\vert \lambda_S\vert^2 }{64 \pi} m_{H'}
\cos^2\theta \left[1-\!\frac{4 m_{\nu_H}^2}{m_{H'}^2}
\right]^{3/2}\!\!\!\!,~
\eea%
where $a_{2}$ is the mixing between light and heavy neutrinos as
defined in Eq.(\ref{nul}), which is of order $0.04$. Thus, for
$m_H \sim 1$ TeV, the decay width $\Gamma(H'\to \nu_l \nu_H) \sim
10^{-3}$. This should be compared with the dominant
decay channel: $H'\to W W$, which has an order one decay width:%
\be%
\Gamma (H' \to W^+ W^-) =  \frac{M_{H'}^3}{16 \pi v^2} \sin^2
\theta \left[1- \frac{4
m_{W}^2}{m_{H'}^2} \right]^{3/2}.%
\ee%
%
\begin{figure}[t]
\begin{center}
\epsfig{file=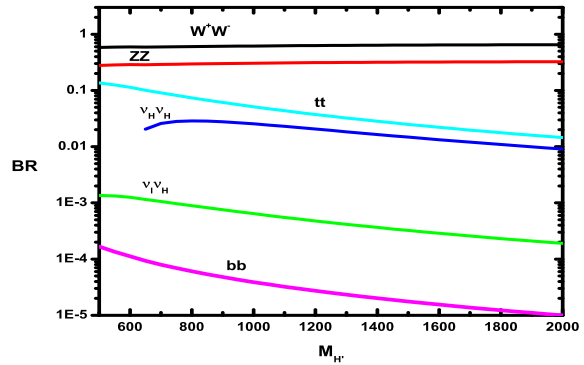, width=8.5cm, height=5.5cm, angle=0}
\end{center}
\vskip -0.5cm \caption{Branching ratios of $H' \to f \bar{f}$ as
function of $M_{H'}$.} \label{fig:BR2}
\end{figure}
The decay branching ratios of $H'$ into $W^+ W^-$, $Z Z$, $\nu_l
\nu_H$, $\nu_H \nu_H$, $t\bar{t}$ and $b\bar{b}$ are shown in
Fig.\ref{fig:BR2} as function of $M_{H'}$. From this figure, it is
clear that the decay of $H'$ is dominated by the same channel of
the SM-like Higgs.  Therefore, these decay channels are
experimentally challenged, due to a large background from the SM
Higgs decays and can not be considered for probing $H'$ at the
LHC. Furthermore, the $H'$ decay into two heavy neutrinos gives
the same signal of two dileptons and missing energy as in $Z'$
decay, but with a smaller cross section. Therefore, the $H'$
production and decay via $H' \to \nu_l \nu_H \to l^+ l^- + {\rm
missing ~ energy}$, as shown in Fig.\ref{fig1}, remains as a
distinctive signal at the LHC that is nearly background free.

The total cross section of this process: $\sigma_{2l}=\sigma(p p
\rightarrow H' \rightarrow \nu_l \nu_H \to l^- l^- + {\rm missing~
energy})$ can be written as %
\be%
\sigma_{2l} \simeq \sigma(p p \rightarrow \tilde{\nu}_{l}
\tilde{\nu}_{H}) \times BR(\nu_H \to l^- W^+) \times BR(W^+ \to
l^+ \nu_l), %
\ee %
where $BR(W^+ \to l^+ \nu_l) \sim 0.1$ and $BR(\nu_H \to l^- W^+)
\sim {\cal O}(1)$, since $\nu_H \to l^- W^+$ is the dominant decay
channel for the heavy neutrino to the SM particles. Finally the
cross section $\sigma(p p \to H' \to \nu_l \nu_H)$ can be
approximated as $\sigma(p p \to H') \times BR(H' \to \nu_l
\nu_H)$, where the $H'$ production is dominated by gluon-gluon
fusion mechanism as shown in Fig.\ref{fig1}. In this case, $\sigma(p p \to H') \sim {\cal O}(0.01)$
as emphasized in Ref.\cite{Emam:2007dy}.
Also from Fig.\ref{fig:BR2}, one can notice that $BR(H' \to \nu_l
\nu_H) \sim 10^{-3}$. Therefore, $\sigma(p p \to H' \to \nu_l
\nu_H) \sim 10^{-5}$. In this case, the total cross section of the
two dilepton signal, which provide indisputable evidence for
probing the $B-L$ extra
Higgs $H'$, is give by %
\bea%
\sigma_{2l} &=&  \sigma(p p\rightarrow H' \rightarrow
l^+ l^- + {\rm missing ~ energy})\nn\\
&\simeq& 10^{-7} {\rm GeV}^{-2} \simeq {\cal O}(100) {\rm pb}.%
\eea%
For this value of cross section, the dilepton and missing energy
signal can be probed at the LHC as a a clear hint for $B-L$ extra
Higgs.

It is worth mentioning that if $m_{H'}
> 2 m_{\nu_{R_H}}$, then the decay width $\Gamma(H' \to \nu_{H}
\nu_{H})$ becomes relevant and may be dominant. However, as
mentioned above, this process leads to a signals of two dileptons
with missing energy similar to the decay of $Z' \to \nu_H \nu_H$
but with a smaller cross section. Therefore, this channel is not
the best for probing $H'$ at the LHC.

Finally, let us note that the above mentioned two dilptons and
missing energy ($4l + \slash\!\!\!\!E_T$) and dilpton plus missing
energy ($2l + \slash\!\!\!\!E_T$) final states are mediated by the heavy neutrinos
$\nu_H$, therefore they are also clean signatures for probing $\nu_H$ at the LHC.

In conclusion, we have constructed a modified version of minimal
TeV scale $B-L$ extension of the SM. In this model, the neutrino
masses are generated through the inverse seesaw mechanism
therefore, the neutrino Yukawa coupling is no longer constrained
to be less than $10^{-6}$. Thus, the heavy neutrinos associated with this model
can be quite feasible at the LHC. We have discussed the main
phenomenological features of this class of models. We showed that
searching for the $Z'_{B-L}$ and heavy neutrinos is accessible via
$4l + \slash\!\!\!\!E_T$ final state, while searching for the
extra Higgs and also heavy neutrino can be accessible through $2l
+ \slash\!\!\!\!E_T$  final state. These final states are very
clean signals at LHC, with negligibly small SM background.

\begin{figure}[b]
\begin{center}
\epsfig{file=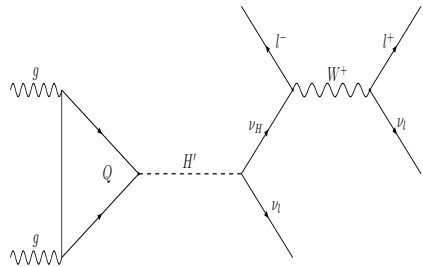, width=5.5cm, height=3.5cm, angle=0}
\end{center}
\caption{$H'$ production and decay into dilepton and missing
energy at the LHC.} \label{fig1}
\end{figure}
%
%
{\it Acknowledgments:}
I would like to thank E. Ma for very useful discussion. This work
was partially supported by the Science and Technology Development
Fund (STDF) Project ID 437, the ICTP Project ID 30, and the
Academy of Scientific Research and Technology.

\end{document}